\documentclass{tMOP2e}
\usepackage{graphicx,bm}
\usepackage{color}

\begin{document}

\title{Theoretical analysis of time delays and streaking effects in XUV photoionization}

\author{
Jing Su$^{a \ast}$,
\thanks{$^\ast$Corresponding author. Email: jing.su@colorado.edu \vspace{6pt}}
Hongcheng Ni$^{a}$, Andreas Becker$^{a}$ and Agnieszka Jaron-Becker$^{a}$\\
\vspace{6pt}
$^{a}${\em{JILA and Department of Physics, University of Colorado, Boulder, CO 80309, USA}}\\\vspace{6pt}
\received{\today}}

\maketitle

\begin{abstract}
We apply a recently proposed theoretical concept and numerical approach to obtain time delays in extreme ultraviolet (XUV) 
photoionization of an 
electron in a short- or long-range potential. The results of our numerical simulations on a space-time grid are compared 
to those for the well-known Wigner-Smith time delay and different methods to obtain the latter time delay are reviewed.
We further use our numerical method to analyze the effect of a near-infrared streaking field on the time delay obtained 
in the numerical simulations.
\end{abstract}

\begin{keywords}
time delay; XUV photoionization; streaking; backpropagation
\end{keywords}

\section{Introduction}

Recent experimental observations revealed substantial time delays in photoionization of atoms from different orbitals \cite{schultze10,klunder11,guenot12}. For some of these observations the so-called attosecond streak camera technique \cite{itatani02} has been used to map the desired time 
information onto the momentum or electron spectrum of the photoelectron as a function of the delay between the ionizing 
extreme ultraviolet (XUV) pulse and a streaking near-infrared (NIR) pulse. Theoretical analysis of the experimental observations 
has mainly focused on two aspects recently \cite{schultze10,kheifets10,zhang10,nagele11,ivanov11,moore11,zhang11,kheifets11,ivanov11pra,ivanov12,sukiasyan12,dahlstrom12,spiewanowski12,guenot12,pazourek12,nagele12}. On the one hand it is the calculation of the time delay, which is assumed to being related 
to the so-called Wigner-Smith (WS) time delay \cite{wigner55,smith60}, which is a well-known concept in scattering theory and different methods have been 
proposed for the calculation \cite{schultze10,kheifets10,nagele11,ivanov11,zhang11,kheifets11,ivanov11pra,dahlstrom12,guenot12,pazourek12,nagele12}. On the other hand there is the question how and to which extent the observation of the time 
delay is influenced by the NIR field, which is used for the observation in the experiment \cite{zhang10,nagele11,ivanov11,dahlstrom12,pazourek12,nagele12}.

In order to contribute to the theoretical analysis we have recently proposed a theoretical concept along with a numerical 
approach to determine time delays in XUV photoionization with and without streaking field \cite{su_submitted}. In this approach 
we make use of a fundamental quantum mechanical definition for the time a particle spends in a certain region of a potential. 
The definition is analogous to that used in the derivation of the so-called WS time delay \cite{wigner55,smith60}. Furthermore, we proposed 
a back-propagation method by which the concept can be applied in actual numerical solutions of the time-dependent Schr\"odinger equation (TDSE)
on the grid. Application and first results for XUV photoionization in the short-range Yukawa potential and the long-range Coulomb 
potential have been presented recently \cite{su_submitted}. 

In this paper we extend our studies and analyze the XUV photoionization in further short-range potentials, such as the Hulth\'en potential 
and the Woods-Saxon potential, as well as the combination of a short- and a long-range potential. The results allow us to establish the 
applicability of our numerical approach in more general. Furthermore, we use the results of our numerical calculations for a comparison 
with the results for the WS time delay. This enables us to discuss and review different approaches to obtain the WS 
time delay for short- as well as long-range potentials. Finally, we present an analysis of the impact of a NIR streaking field 
on the numerical results for the time delay in different potentials. Our results show, in general, that the effect is small as long 
as the intensity of the NIR field does not exceed $1\times 10^{13}$ W/cm$^2$.

\section{Numerical simulations of time delays in XUV photoionization}

In this section we discuss the application of our recently proposed numerical
approach \cite{su_submitted} to determine time delays in photoionization of an electron by
an ultrashort XUV pulse. To this end, we first briefly
outline the theory as well as the different potentials used in the present calculations.
We then present results of test calculations showing
some characteristic features of our time delay calculations for short- and long-range potentials
as well as the combination of both.

\subsection{Outline of theoretical approach}

We define the time delay in XUV photoionization as the difference between the time $t_{\Psi,R}$
an initially bound electron needs to leave a certain region $R$ of a potential due to the interaction with the XUV field
and the time $t_{\Psi^{(0)},R}$ a free particle spends in the same region \cite{su_submitted}
(Hartree atomic units, $e = m = \hbar = 1$ are used throughout the paper):
\begin{equation}
\Delta t_{\Psi,R} = t_{\Psi,R} - t_{\Psi^{(0)},R},
\label{num_delay}
\end{equation}
with
\begin{equation}
t_{\Psi,R} = \int_{-\infty}^{\infty}dt \int_R d{\bm r}
\frac{|\Psi^{(ion)}({\bm r},t)|^2}{\int_{-\infty}^{\infty}d{\bm r}|\Psi^{(ion)}({\bm r},t\rightarrow\infty)|^2},
\label{time_ion}
\end{equation}
and
\begin{equation}
t_{\Psi^{(0)},R} = \int_{-\infty}^{\infty}dt \int_R d{\bm r} |\Psi^{(0)}({\bm r},t)|^2.
\label{time_free}
\end{equation}
$\Psi$ is the initial bound state of the electron in the potential $V({\bf r})$ and $\Psi^{(0)}$ is the free state of the
electron corresponding to the ionizing part of the wavefunction $\Psi^{(ion)}$
after transition from the initial state to the continuum by absorbing an XUV photon.
In Eqs.\ (\ref{time_ion}) and (\ref{time_free}) the wavefunctions $\Psi^{(ion)}$ and $\Psi^{(0)}$ are renormalized in order to get physically reasonable times.

The definition given in Eq.\ (\ref{num_delay}) is based on the well-known concept of a time delay in particle scattering of a
potential \cite{wigner55,smith60}. As in these early studies in scattering the time delay $\Delta t_{\Psi,R}$ has a finite
limit as the radius of $R$ grows to infinity for short-range potentials $V({\bf r})$. In contrast, for long-range potentials
such as the Coulomb potential $\Delta t_{\Psi,R}$ diverges in this limit \cite{su_submitted}. Furthermore, according
to the definition above the time delays are
negative for propagation of the particle in an attractive potential $V({\bf r})$, since a free wave packet spends more time in a
given region $R$ than the corresponding wave packet that has the same asymptotic energy propagating in the attractive potential.

In order to use the above definition in a numerical simulation of an XUV photoionization process we use the back-propagation technique \cite{su_submitted}.
To this end, we first solve the corresponding TDSE of the system,
initially in the state $\Psi({\bm r},t=0)$, under the interaction with the external light field on a space-time grid:
\begin{equation}
i\frac{\partial}{\partial t} \Psi({\bm r},t)
=
\left(\frac{{\bm p}^2}{2} + V({\bm r}) + V_{\text{light}}(t)
\right)
\Psi({\bm r},t),
\label{schroedinger}
\end{equation}
where ${\bm p}$ is the momentum operator and $V_{\text{light}}(t)$ represents the interaction with the ionizing light field,
while $V({\bm r})$ includes the (short- or long-range) electrostatic potential as well as, if used in the calculations, a streaking-field
$V_\text{streak}$. We proceed with the (forward) propagation to large distances on the grid at least until the interaction with the
ionizing light field as well as the interaction with the streaking field ceases. This allows us to spatially separate the
ionizing part of the wavefunction from the remaining bound parts. Alternatively, we could project onto analytically or numerically known bound
states for the separation. Then we propagate the remaining ionizing part of the wavefunction backwards in time
without taking account of the interaction with the light field $V_{\text{light}}$ once within the potential $V({\bm r})$:
\begin{equation}
i\frac{\partial}{\partial t} \Psi^{(ion)}({\bm r},t)
=
\left(\frac{{\bm p}^2}{2} + V({\bm r})
\right)
\Psi^{(ion)}({\bm r},t)
\end{equation}
and once as a free particle:
\begin{equation}
i\frac{\partial}{\partial t} \Psi^{(0)}({\bm r},t)
=
\frac{{\bm p}^2}{2}
\Psi^{(0)}({\bm r},t)
\end{equation}
The corresponding times $t_{\Psi,R}$ and $t_{\Psi^{(0)},R}$ can then be calculated for regions $R$ with radius smaller than
the location of the wave packet at the start of the back-propagation.

\subsection{Model systems and numerical simulation}

Below we present numerical results for the application of this theoretical method to the photoionization of an electron initially bound in several
different model potentials, namely the long-range Coulomb potential in 1D:
\begin{equation}
V_{\text{C}}(x)=-\frac{Z}{\sqrt{x^2+a_1}},
\label{Coulomb}
\end{equation}
with $Z=3.0$ is the effective nuclear charge, and $a_1=2.0$ is the soft-core parameter; short-range interactions represented by
the 1D Hulth\'en potential $V_{\text{H}}(x)$ and the 1D Woods-Saxon potential $V_{\text{WS}}(x)$, given by
(for studies using the short-range Yukawa potential, see \cite{su_submitted}):
\begin{equation}
V_{\text{H,WS}}(x)=\frac{V_0}{q-e^{\sqrt{x^2+a_2}/b_2}},
\label{Hulthen-WoodsSaxon}
\end{equation}
with $V_0=0.5$, $q=1.0$, $a_2=1.5$, and $b_2=6.0$ for the Hulth\'en potential and
$V_0=5.0$, $q=-1.0$, $a_2=1.85$, and $b_2=3.0$ for the Woods-Saxon potential, and a combination of $V_{\text{C}}(x)$
and $V_{\text{WS}}(x)$:
\begin{equation}
V_{\text{C-WS}}(x)=-\frac{Z}{\sqrt{x^2+a_1}}\frac{1}{1+e^{(|x|-x_0)/b_1}}.
\label{Coulomb-WoodsSaxon}
\end{equation}
with $x_0=30.0$ or $100.0$, and $b_1=1.0$ for the combined potential.
For these potential parameters, we obtained almost identical ground state energies for the potentials, namely
$-1.7118$ (Coulomb potential $V_{\text{C}}(x)$),
$-1.7088$ (Hulth\'en potential $V_{\text{H}}(x)$),
$-1.7100$ (Woods-Saxon potential $V_{\text{WS}}(x)$) ,
and $-1.7118$ (combined Coulomb-Woods-Saxon potential $V_{\text{C-WS}}(x)$).
We note that the combined potential has the same ground state energy as the Coulomb potential, since the
Woods-Saxon factor in Eq. (\ref{Coulomb-WoodsSaxon}) represents a smoothed step-function, which does not influence
the short-range part of the Coulomb potential.

For the interaction with the XUV light pulse we used length gauge, i.e.
\begin{equation}
V_{\text{light}}(t) = E_{\text{XUV}}(t) x
\end{equation}
where $E_{\text{XUV}}$ represents a linearly polarized pulse with a sin$^2$ envelope, i.e.,
\begin{equation}
E_{\text{XUV}}(t)=E_0 \sin^2(\pi t/\tau) \sin(\omega t+\phi),
\label{XUV_pulse}
\end{equation}
where $E_0$ is the peak amplitude, $\tau$ is the pulse duration, $\omega$ is the central frequency,
and $\phi$ is the carrier-envelope phase (CEP). The streaking field $V_\text{streak}$, if used in the calculations,
is represented in an analogous form.

To solve the corresponding TDSE in our simulations, we used the Crank-Nicolson method in a grid representation, with, in general,
a spatial step of $\delta x =0.1$ and a time step of $\delta t=0.02$. In case of the short-range potentials (incl.\ the combined
Coulomb-Woods-Saxon potential), the grid
extended from $-9000$ a.u.\ to $9000$ a.u., while a larger grid of $-14000$ a.u.\ to $14000$ a.u.\ was adopted to forward
propagate the ionizing wave packet for the Coulomb potential. The larger grid in case of the long-range potential assures that
the final momentum of the electron at the end of the forward propagation is close to the asymptotic one, which reduces the error
in the calculation of the time the free particle spends in the potential \cite{su_submitted}.

The initial ground states were obtained by imaginary time propagation. For the back propagation we propagated the
ionizing parts at negative and positive $x$ backwards in time independently, either under the influence of the potential,
or as a free particle. We determined the corresponding times $t_{\Psi,R}$ and $t_{\Psi^{(0)},R}$ for both parts of
the ionizing wavefunction and added the two contributions. In the 1D calculations we defined the region as
$R = [\pm x_\text{inner}, \pm x_\text{outer}]$, where $x_\text{inner}$ and $x_\text{outer}$ are the inner and outer boundaries,
respectively, and the $\pm$-signs apply to back-propagation of the two parts of the ionizing wave packet along the
positive/negative $x$-axis, respectively. We absorbed the wavefunction beyond the inner boundary $x_\text{inner}$
using the exterior complex scaling method \cite{ho83,mccurdy04}.

\subsection{Some characteristic features}

\begin{figure}[t]
  \begin{center}
  \includegraphics[scale=0.45]{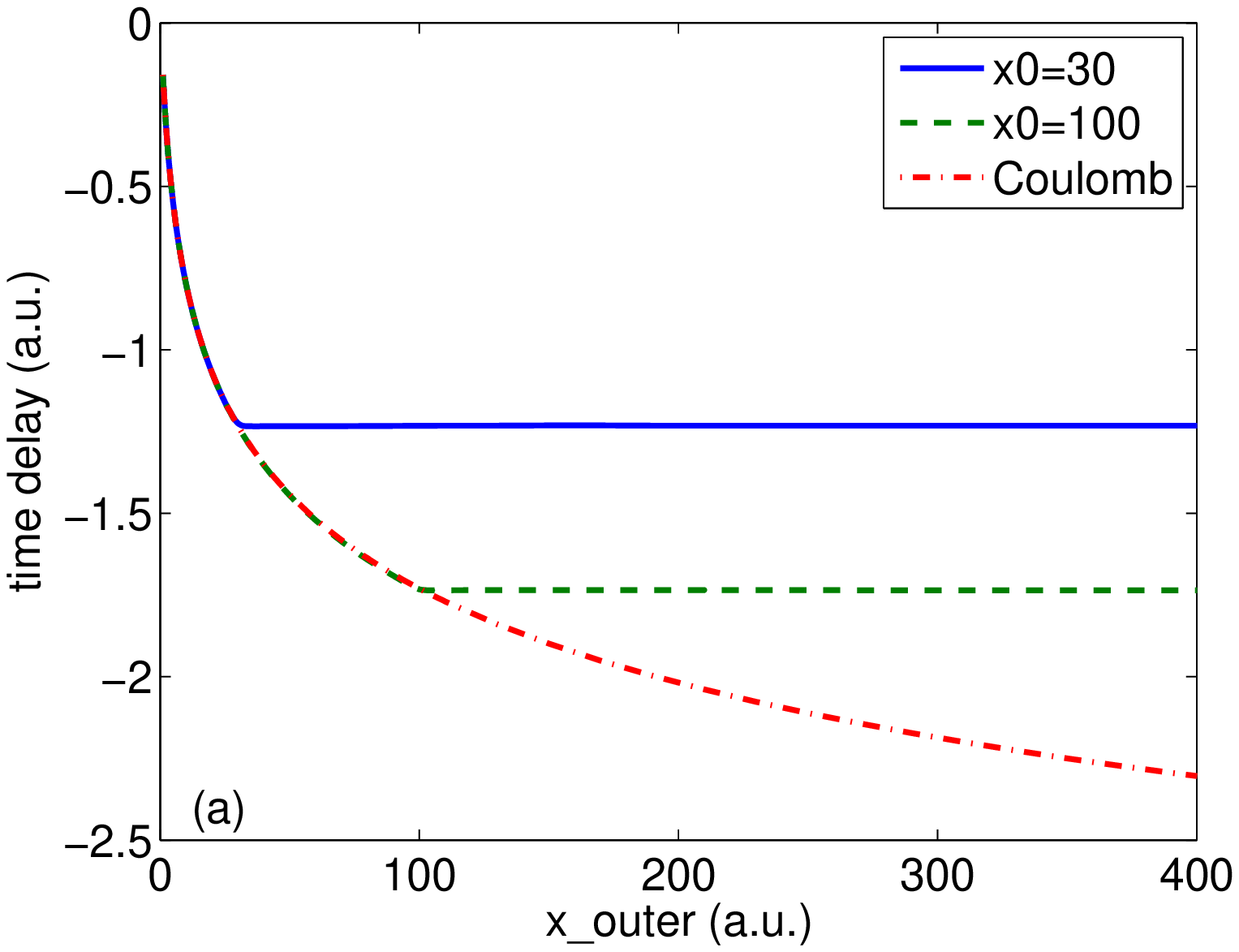}
  \includegraphics[scale=0.45]{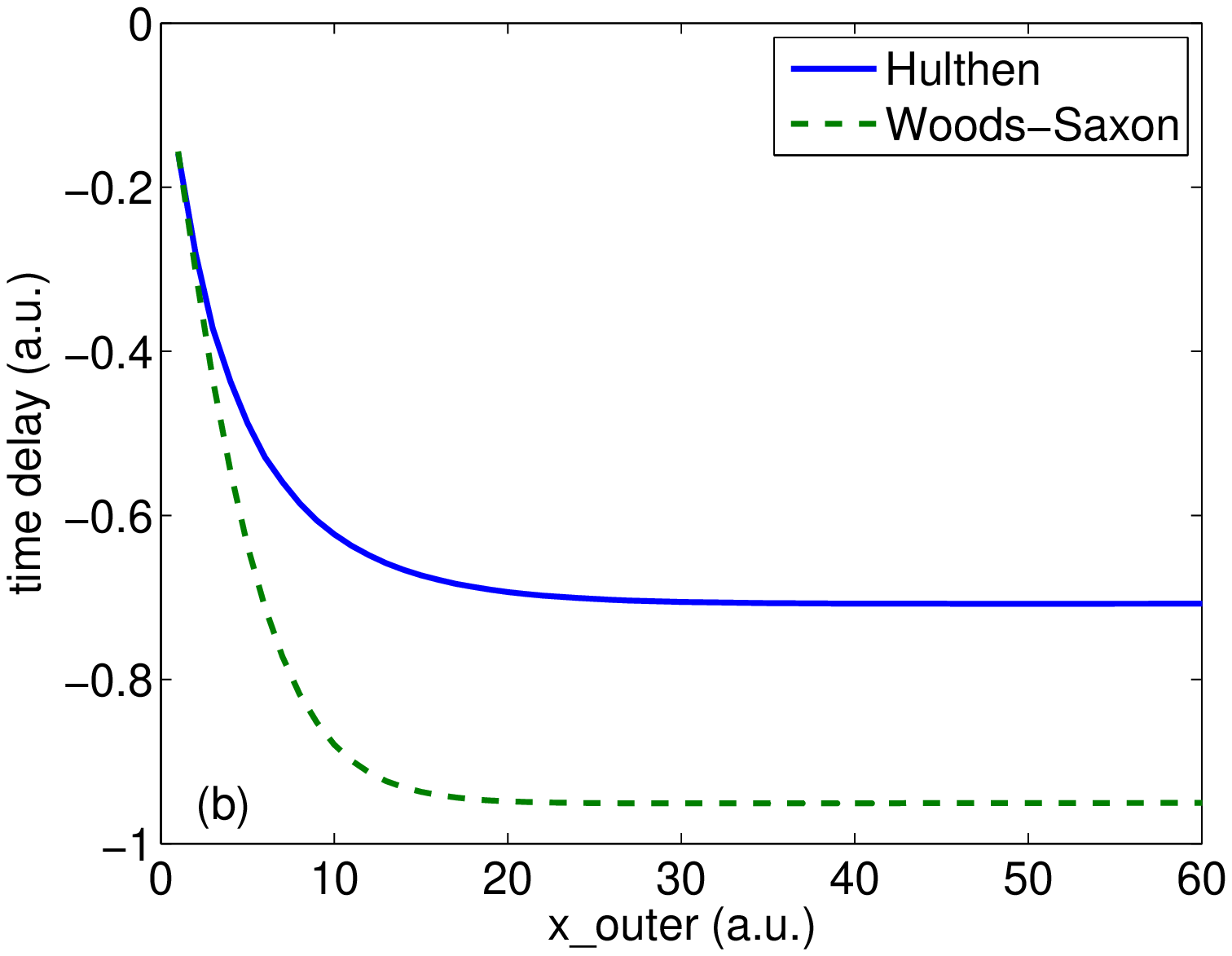}
  \end{center}
  \caption{
Time delays as a function of the outer integration boundary $x_\text{outer}$ for different potentials:
(a) Coulomb potential (red dash-dotted line) and combined Coulomb-Woods-Saxon potential (blue solid line for $x_0=30$, green dashed line for $x_0=100$), (b) Hulth\'en potential (blue solid line) and Woods-Saxon potential (green dashed line). In all calculations we have used an XUV pulse with peak intensity $I=1\times10^{15}$ W/cm$^2$, central frequency $\omega=100$ eV, pulse duration $\tau=400$ as, and carrier-envelope phase $\phi=0$ for the ionization. (The color version of this figure is included in the online version of the journal.)
}
\label{delay_outer_boundary}
\end{figure}

\begin{figure}[t]
  \begin{center}
  \includegraphics[scale=0.45]{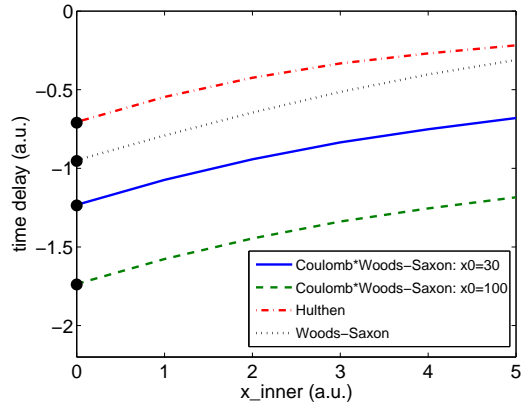}
  \end{center}
  \caption{Time delays as a function of inner integration boundary $x_{\text{inner}}$. Three short-range potentials are considered here: Hulth\'en potential (red dash dotted line), Woods-Saxon potential (black dotted line), and the combined Coulomb-Woods-Saxon potential (blue solid line for $x_0=30$ and green dashed line for $x_0=100$). Laser parameters are the same as in Fig. \ref{delay_outer_boundary}. The corresponding WS time delays were shown as black circles. (The color version of this figure is included in the online version of the journal.)}
\label{delay_inner_boundary}
\end{figure}

In Figs.\ \ref{delay_outer_boundary} and \ref{delay_inner_boundary} we present some characteristic results obtained for the
time delay $\Delta t_{\Psi,R}$ in the different potentials without streaking field.
As can be seen from the results in Fig.\ \ref{delay_outer_boundary}, $\Delta t_{\Psi,R}$ depends on the size of the region $R$ and
is negative in all cases studied. It converges to a finite limit as $R$ increases towards infinity for short range potentials only.
These results are
in agreement with the general features of a time delay in scattering theory \cite{wigner55,smith60}.
In the case of the combined Coulomb-Woods-Saxon potential
(blue solid and green dashed lines in Fig.\ \ref{delay_outer_boundary}(a)), one can most clearly see that the time delay converges
for distances beyond the effective range of the potential.
Our results for the Coulomb potential (red dash-dotted curve
in Fig.\ \ref{delay_outer_boundary}(a)) also confirm the expectation that there is no limit for the time delay
as the outer boundary increases in such a long-range potential. This is due to the well-known logarithmic divergence related with
ionization from any bound state within the potential. The effect of the long-range tail of the Coulomb potential
can be clearly seen from the comparison of corresponding results
(red dash-dotted line) with those for the combined Coulomb-Woods-Saxon potentials
in Fig. \ref{delay_outer_boundary}(a). We however note that for any finite distance the
time delay is a well-defined finite number for each of the potentials. This allows us e.g.\ to study
effects of an ultrashort streaking pulse even for the long-range Coulomb potential with the present theoretical approach
\cite{su_submitted}.

We also note from the results in Fig.\ \ref{delay_outer_boundary} that in each case the absolute value of the time delay increases
most strongly in the region close to the center of the potential, where the potential is strongest.
Therefore, our theoretical results strongly depend on the choice of the inner boundary $x_\text{inner}$ of the region $R$,
which can be seen from the results in Fig.\ \ref{delay_inner_boundary}. In the latter calculations we
fixed the outer boundary of $R$ at $x_\text{outer}=500$, which is large enough to obtain converged results
for all short-range potentials considered here.
In our studies presented below
we have chosen $x_\text{inner}=0$, which corresponds to the expectation value of $x$ for all the bound states considered here.
As indicated in Fig.\ \ref{delay_inner_boundary} and discussed below, for this choice of $x_\text{inner}$ our results
are also in excellent agreement with those for the so-called WS time delay, if applicable (black circles in Fig.\ \ref{delay_inner_boundary}).

\section{Numerical results and discussion}

In this section we first present comparisons of our numerical results for time delays with those for the
so-called WS time delay for short-range potentials. Then, we use the fact that all time delays are finite
for any finite region $R$ to study the effect of a NIR laser pulse, as used
in recent streaking experiments \cite{schultze10}, on the time delays.

\subsection{Comparison with Wigner-Smith time delays}

For particle scattering of a potential the time delay, defined analogous to the expression given in Eq.\ (\ref{num_delay}),
can be given in the limit of an infinitely large region $R$ as \cite{wigner55}:
\begin{equation}
\Delta t_{\text{WS}}= \frac{d\varphi}{dE},
\label{WS_delay}
\end{equation}
where $\varphi$ is the phase shift induced by the potential $V({\bf r})$ upon scattering of the particle.
This expression is also commonly known as the (asymptotic) WS time delay \cite{wigner55,smith60} and is well defined
for any short-range potential. In contrast, application of the concept and formula in case of a long-range potential
is questionable, since the required limit of the time delay does not exist (c.f., e.g., \cite{smith60}).

We therefore restrict ourselves to a comparison of our numerical results for the time delay with those
for the WS time delay in the case of short-range potentials. We obtained the WS time
delay for photoionization in two ways. On the one hand we made use of Eq.\ (\ref{WS_delay}) and determined the
phase derivative in a time-independent scattering approach. To this end, we first considered scattering of an electron,
incident from $x = -\infty$ with a momentum $k$ of the short-range potential of interest. We solved the corresponding
time-independent Schr\"odinger equation numerically using the fourth order Runge-Kutta method up to $|x| = 500$, i.e.\
well beyond the effective range of any short-range potential considered here.
Then we projected the numerical solution onto the appropriate plane wave solutions for $x \rightarrow \pm\infty$, and
obtained the WS time delay for the scattering process $\Delta t_{\text{WS}}^{\text{(scat)}}$ as the derivative
of the phase shift in the plane wave propagating in positive $x$-direction with respect to the energy of the incident particle.
For a specific XUV photoionization process we averaged the results over the energy spectrum of the wave packet, as
obtained in our time-dependent numerical simulations. Finally, we divided the result by two
to take account of the fact that photoionization can be considered as a half-scattering process.

\begin{figure}[t]
  \begin{center}
  \includegraphics[scale=0.45]{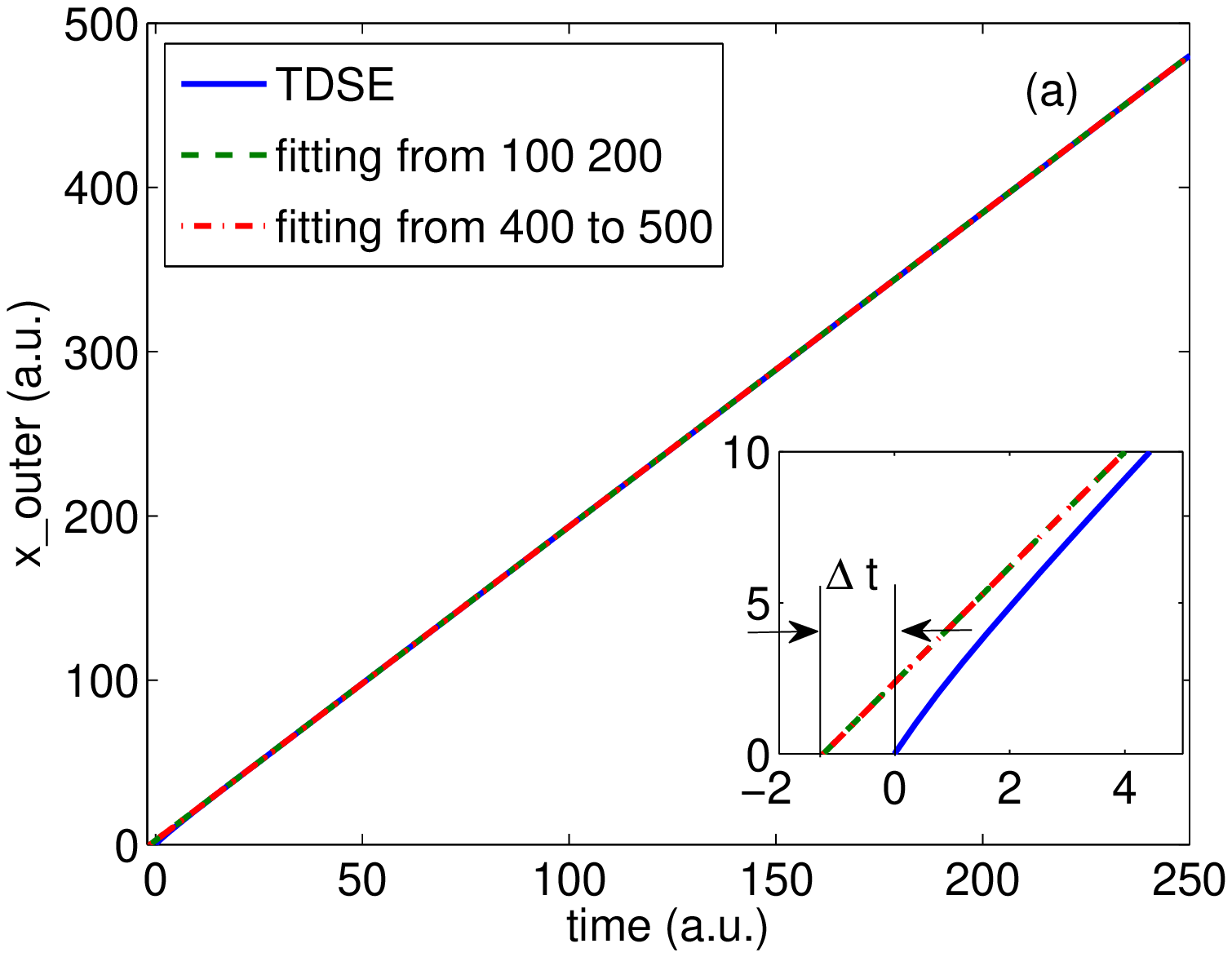}
  \includegraphics[scale=0.45]{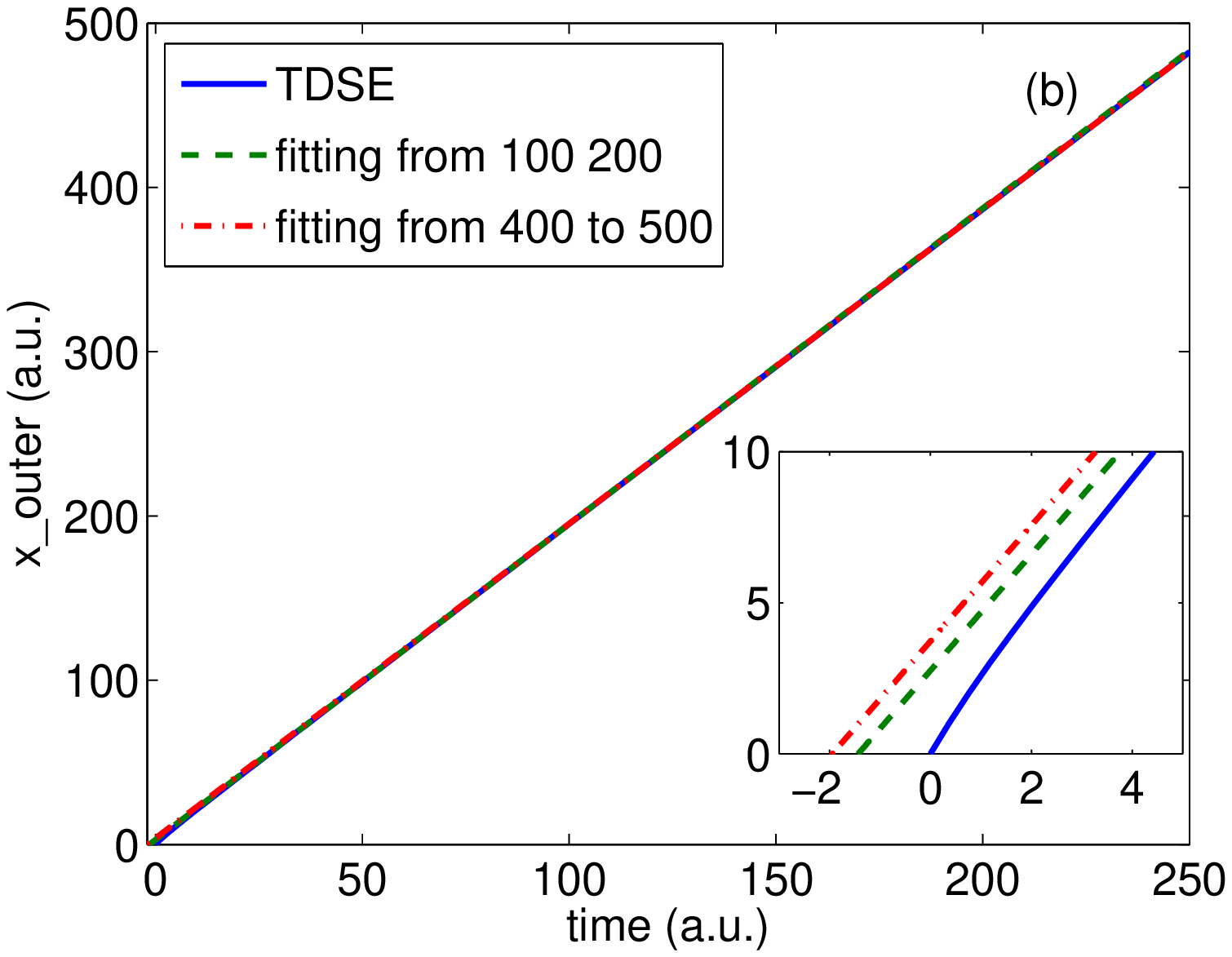}
  \end{center}
  \caption{Extrapolation method to calculate the WS time delay. The solid blue lines show the outer integration boundary $x_\text{outer}$ as a function of time for (a) the combined Coulomb-Woods-Saxon potential with $x_0=30$ and (b) the Coulomb potential. The green dashed and red dash-dotted lines are obtained by linearly fitting the solid blue lines in two regions: $100$ to $200$ and $400$ to $500$ respectively. The two insets show the behaviors near $t=0$ for each line. $\Delta t$ in panel (a) corresponds to the WS time delay for a short-range potential. (The color version of this figure is included in the online version of the journal.)}
\label{delay_extrapolate}
\end{figure}

We also employed an alternative concept to determine the WS time delay, which has been used in several 
recent theoretical works \cite{schultze10,kheifets10,zhang11}.
It is based on the assumption that - significantly far away from the center of a short-range potential - the trajectory of the
electron depends linearly on the propagation time. The time delay is then given as the time difference between 
the time obtained 
by extrapolating the linear part of the electron trajectory back to the center of the potential and the time zero. 
Usually, the electron trajectory is calculated as the expectation value of the ionizing wave packet as a function of time. 
Here, we instead obtained the trajectory by plotting the outer boundary $x_\text{outer}$ of the region $R$ as a function of 
$t_{\Psi_{i},R}$ (blue solid lines in Fig. \ref{delay_extrapolate}). In case of a short-range potential the trajectory 
beyond the effective range of the potential was then fitted to a linear line. As an example, we present in 
Fig. \ref{delay_extrapolate}(a) the results for the combined Coulomb-Woods-Saxon potential with $x_0=30$. It can be seen that the extrapolation does not depend on the 
range over which the trajectory is fitted (cf., green dashed and red dash-dotted lines) and the WS time delay 
can be determined from the intersection of the extrapolated curves with the base line (see inset of 
Fig.\ \ref{delay_extrapolate}(a)). In contrast, this extrapolation method is not applicable in the case of Coulomb potential. 
Due to the logarithmic divergence in the trajectory a fit to a linear line is not possible. 
Linear fits to different parts of the trajectory indeed lead to a significant variation in the extrapolated time delay
(see green dashed and red dash-dotted lines in Fig. \ref{delay_extrapolate}(b)). This is in agreement with the 
fact that the (asymptotic) WS time delay is an ill-defined concept for a long-range potential such as the 
Coulomb potential.

\begin{figure}[t]
  \begin{center}
  \includegraphics[scale=0.40]{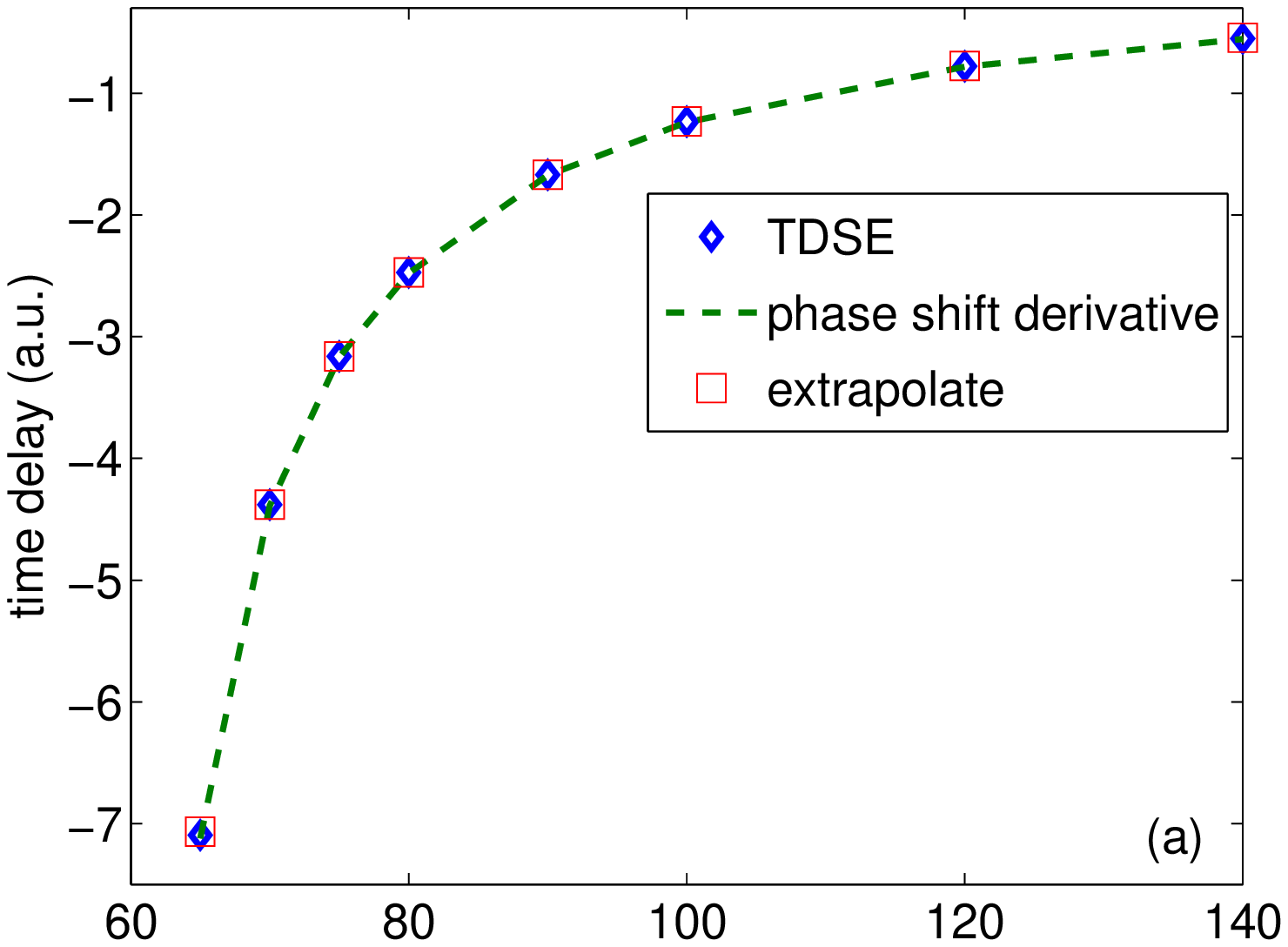}
  \includegraphics[scale=0.40]{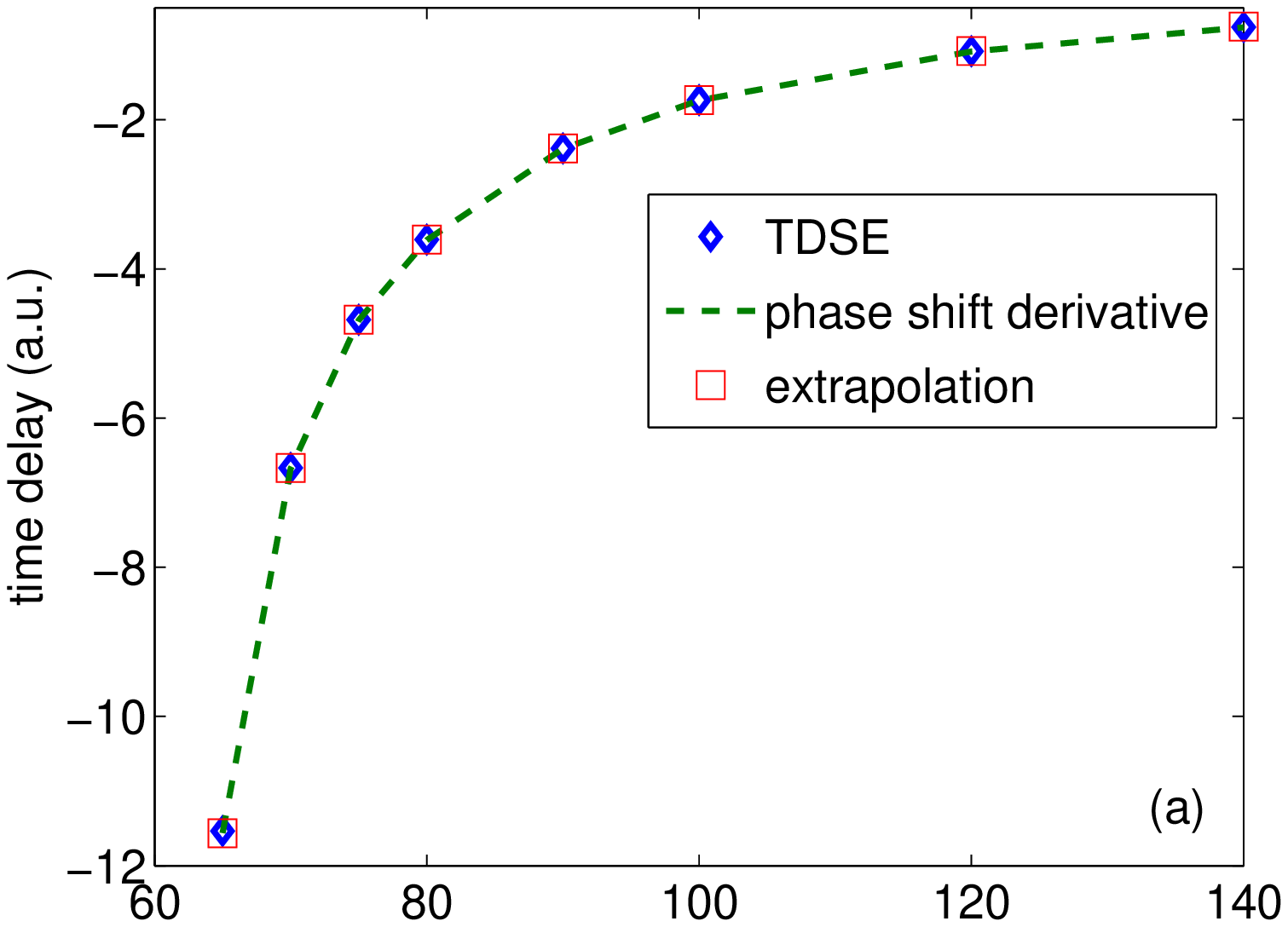}\\
  \includegraphics[scale=0.40]{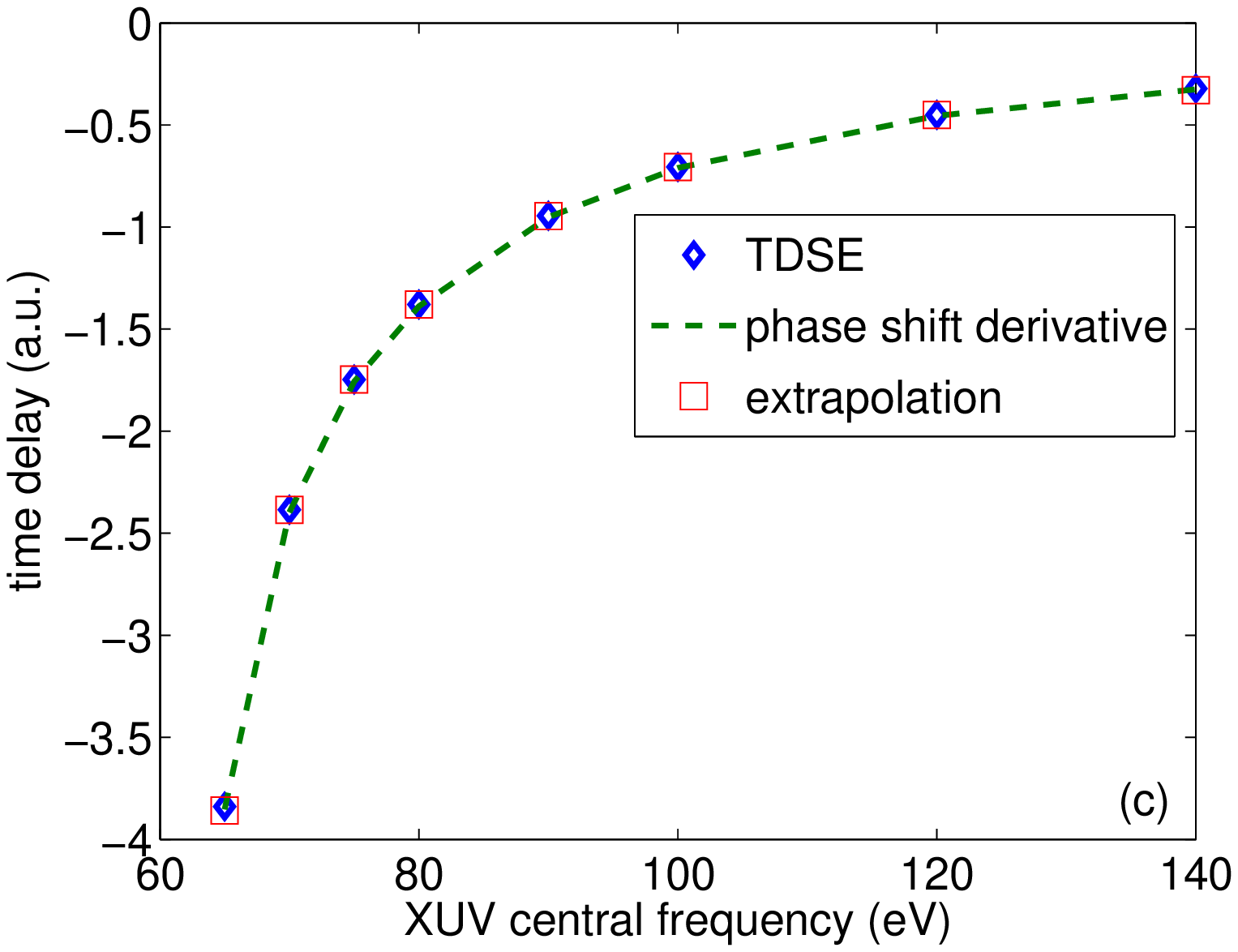}
  \includegraphics[scale=0.40]{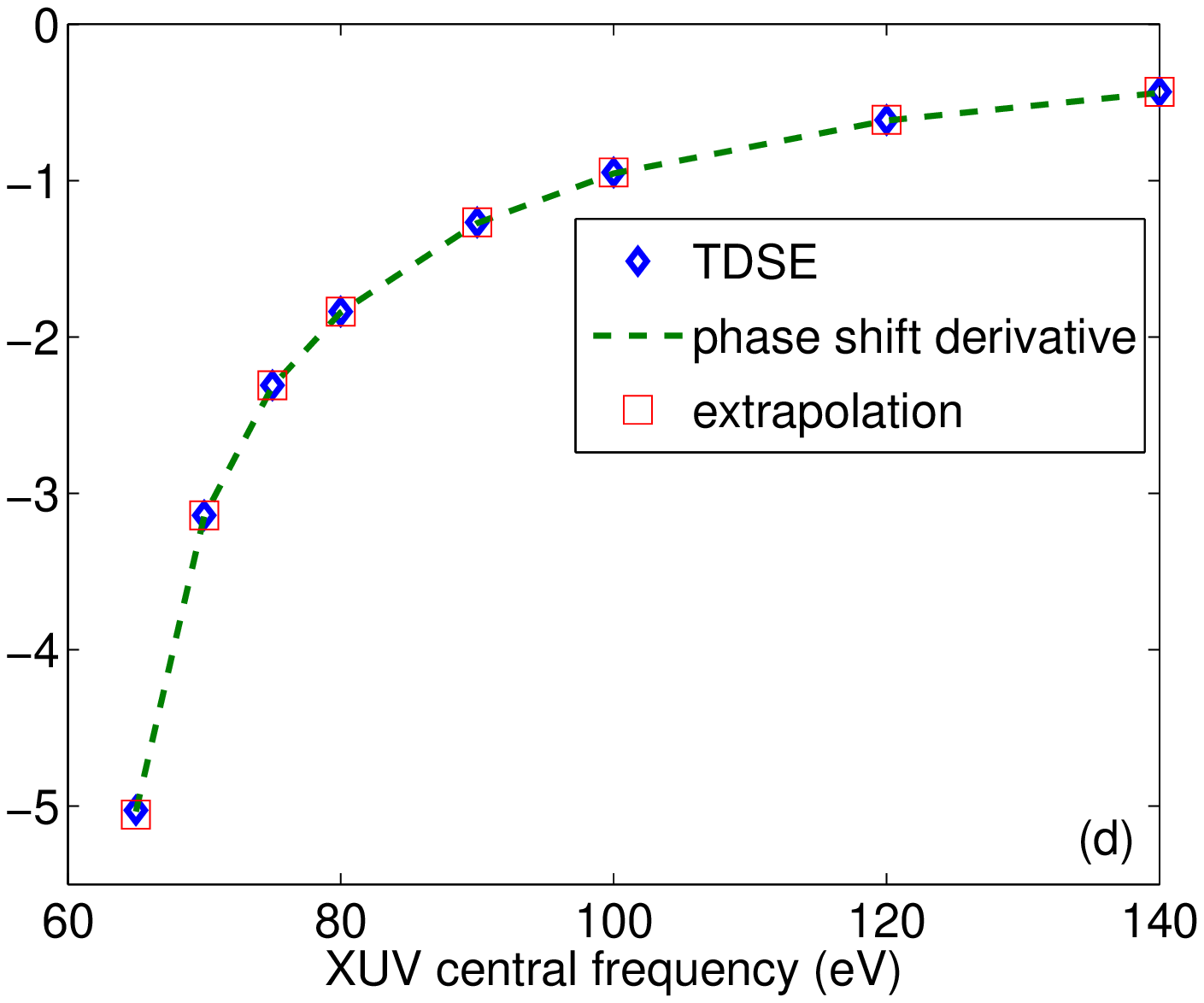}
  \end{center}
  \caption{
Time delays as a function of XUV central frequency $\omega$: (a) combined Coulomb-Woods-Saxon potential with $x_0=30$, (b) combined Coulomb-Woods-Saxon potential with $x_0=100$, (c) Hulth\'en potential, and (d) Woods-Saxon potential. In each panel, three sets of time delays are shown: TDSE results (blue diamonds), WS delays using the extrapolation method (red open squares), and WS delays obtained via the phase shift derivative (dashed lines). Other laser parameters are: $I=1\times10^{15}$ W/cm$^2$, $\tau=400$ as, and $\phi=0$.
}
\label{delay_omega}
\end{figure}

\begin{figure}[t]
  \begin{center}
  \includegraphics[scale=0.40]{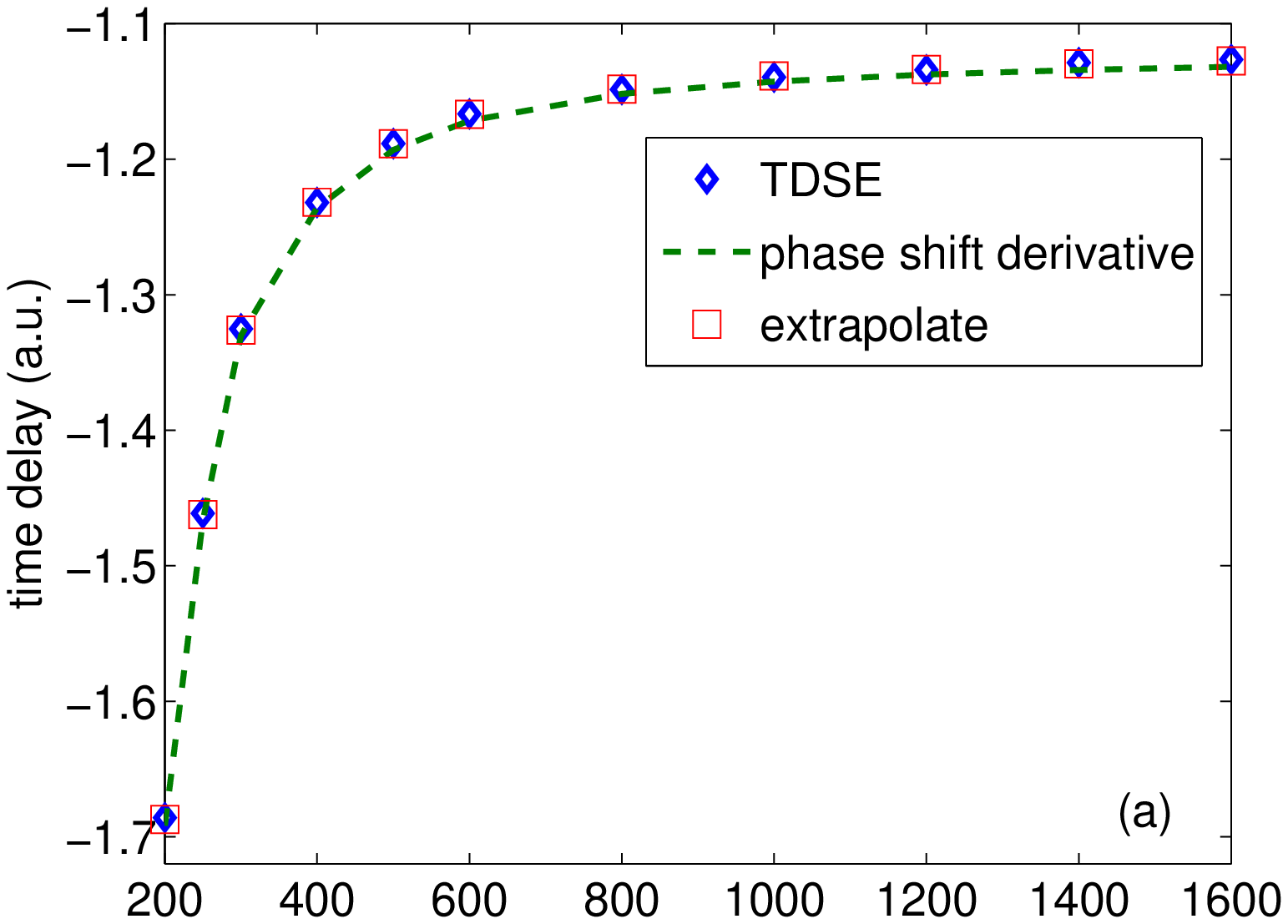}
  \includegraphics[scale=0.40]{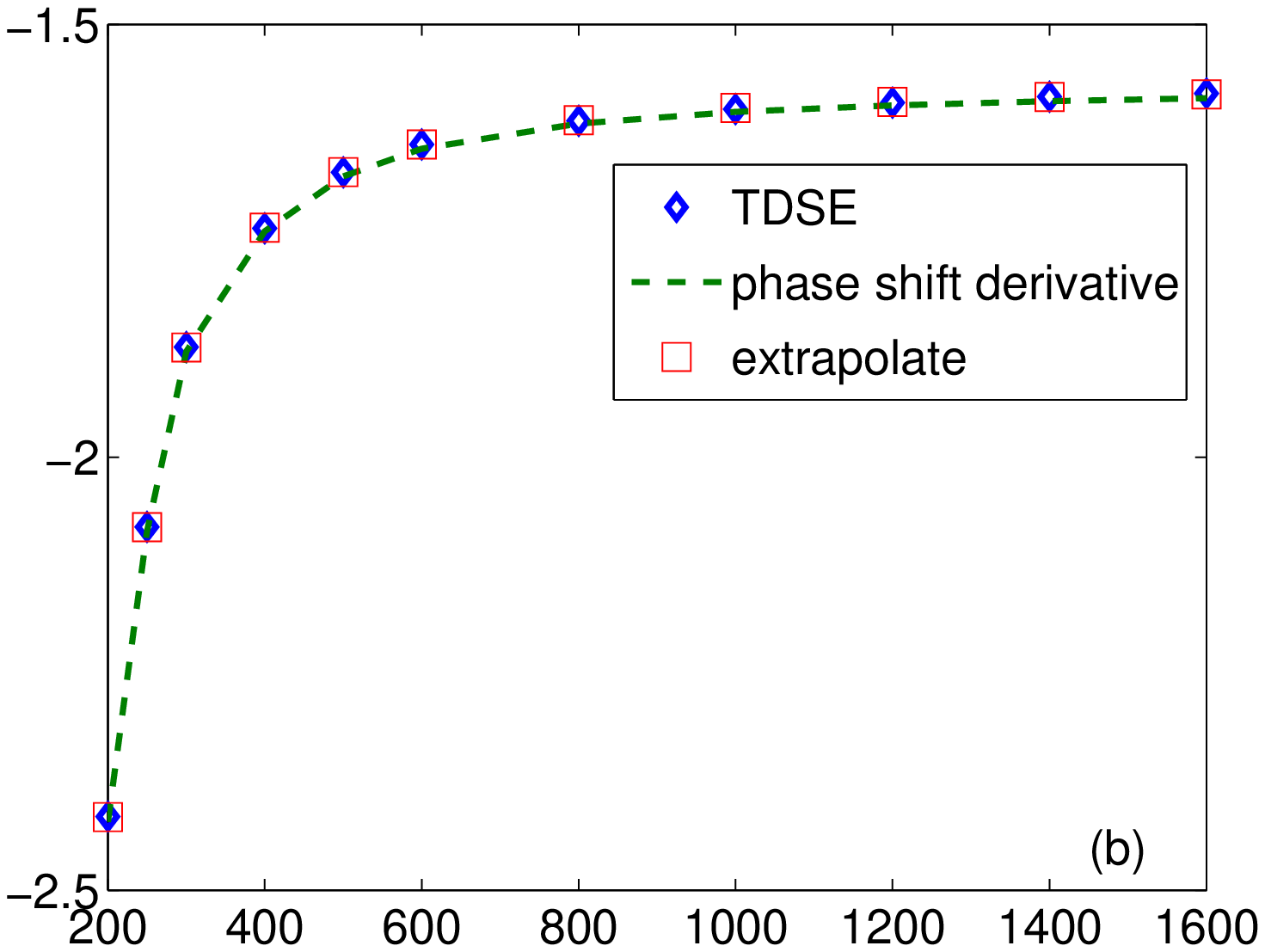}\\
  \includegraphics[scale=0.40]{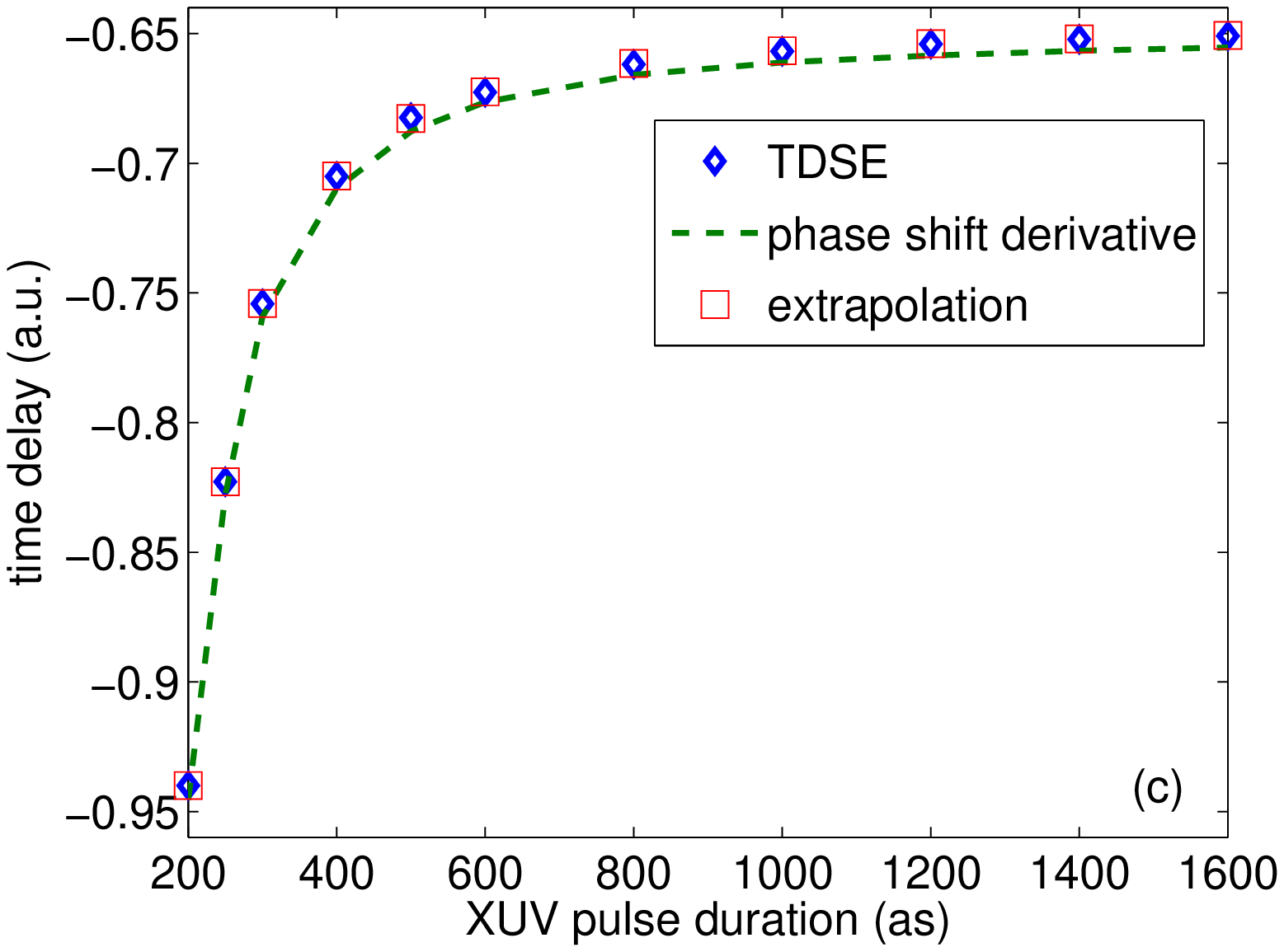}
  \includegraphics[scale=0.40]{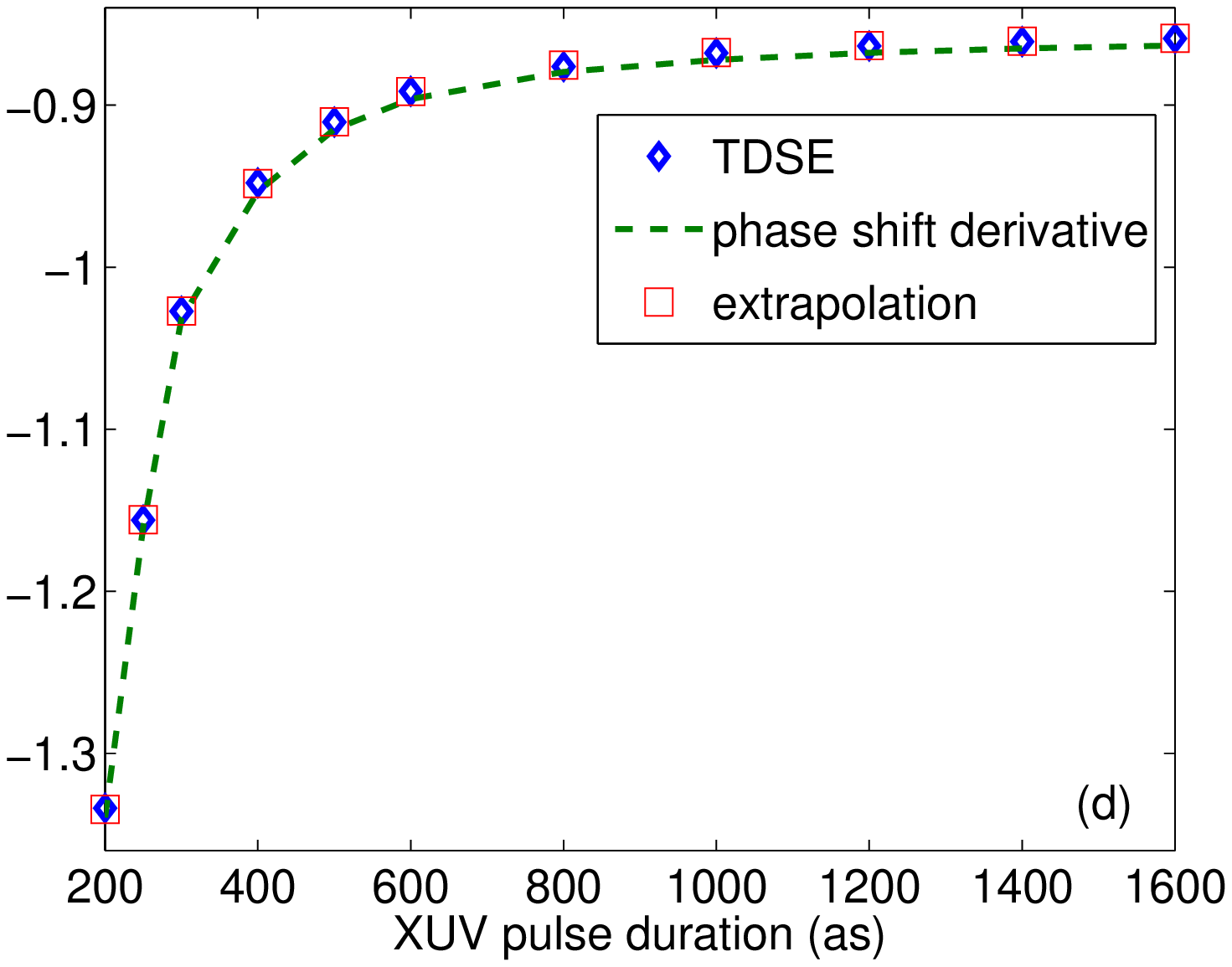}
  \end{center}
  \caption{ Time delays as a function of XUV pulse duration $\tau$. All the symbols have the same meanings as in Fig.\ \ref{delay_omega}. Other laser parameters were: $I=1\times10^{15}$ W/cm$^2$, $\omega=100$ eV and $\phi=0$. (The color version of this figure is included in the online version of the journal.)
}
\label{delay_pulseduration}
\end{figure}

Our numerical results are in excellent agreement with those for the WS time delay in case of the short-range 
potentials. For example, in Fig.\ \ref{delay_inner_boundary} the WS time delays are shown as black circles and are 
found to agree well with the numerical results for $x_\text{inner} = 0$. Next, we compare in Figs.\ \ref{delay_omega} and 
\ref{delay_pulseduration} our numerical results (blue diamonds) for the time delay as a function of XUV frequency at a fixed pulse 
duration of $\tau=400$ as and as a function of the XUV pulse duration at a fixed central frequency of $\omega=100$ eV 
with the WS time delays obtained either via the calculation of the energy derivative of the phase shift 
(dashed lines) or the extrapolation method (red open squares). One may notice that there is a small discrepancy between the time delays from the time-dependent simulations and the 
WS time delay calculated as the phase shift derivative from the time-independent approach, which can be most clearly seen 
in Fig.\ \ref{delay_pulseduration}(c). This discrepancy is related to our choices of grid parameters. Test calculations 
have shown that this discrepancy can be removed by further reducing the spatial step size.

The latter results also agree well with qualitative expectations. The absolute value of the time delay decreases towards 
zero as the frequency of the ionizing XUV laser pulse and, hence, the final kinetic energy of the emitted electron increase, 
since the effect of any potential becomes more and more negligible (see panels (a)-(d) in Fig.\ \ref{delay_omega} for the 
different short-range potentials). 
Furthermore the absolute value of the time delay decreases with an increase of the XUV pulse duration 
(see Fig.\ \ref{delay_pulseduration}), which is closely related to the dependence on the XUV frequency. 
Due to the finite pulse duration the time delay obtained for the wave packet can be considered as an average over 
contributions at particular energies within a certain bandwidth (weighted by the ionization probability at a given energy). 
Since the time delay does not change linearly with the kinetic energy, the value obtained for a wave packet will be smaller 
than its contribution at the expectation value of the kinetic energy. This difference decreases and, thus, the time delay 
for the wave packet increases as the energy bandwidth of the wave packet decreases, i.e.\ as the pulse duration increases. 
Besides, the expectation value of the kinetic energy of the ionizing wavepacket increases as the XUV pulse 
duration increases, which also causes the time delay to increase.

\subsection{Streaking effects}

Finally, we study the effect of an additional external NIR field on the time delay obtained in our numerical 
simulations. This 
plays an important role in view of recent observations of time delays using the attosecond 
streaking camera technique \cite{schultze10,itatani02}. In the latter a weak NIR field is used to map time onto momentum and retrieve from the 
momentum (or, energy) spectrum of the photoelectron as a function of the delay between ionizing XUV pulse and 
streaking NIR pulse information about time delays between different processes. 
 
As mentioned above we take account of the streaking field as one of the potential terms in 
$V({\bm r})$ in Eq.\ (\ref{schroedinger}) and therefore consider the streaking field on equal footing with the short- or long-range
electrostatic potential of interest. 
In this way we obtain as result of the back-propagation step the time delay as
\begin{equation}
\Delta t^{(\text{NIR})}_{\Psi,R} = t_{\Psi,R}^\text{(NIR)} - t_{\Psi^{(0)},R},
\label{num_delay_IR}
\end{equation}
where $t_{\Psi,R}^\text{(NIR)}$ is the time the ionizing wavepacket spends in region $R$ in the presence of the 
electrostatic potential and the NIR streaking field. In our calculations we consider XUV photoionization from 
the ground state of the short-range potentials and checked that the ionization induced by the NIR field is negligible 
up to an streaking field intensity of $1\times10^{13}$ W/cm$^2$ in each of the cases presented below.

\begin{figure}[t]
  \begin{center}
  \includegraphics[scale=0.45]{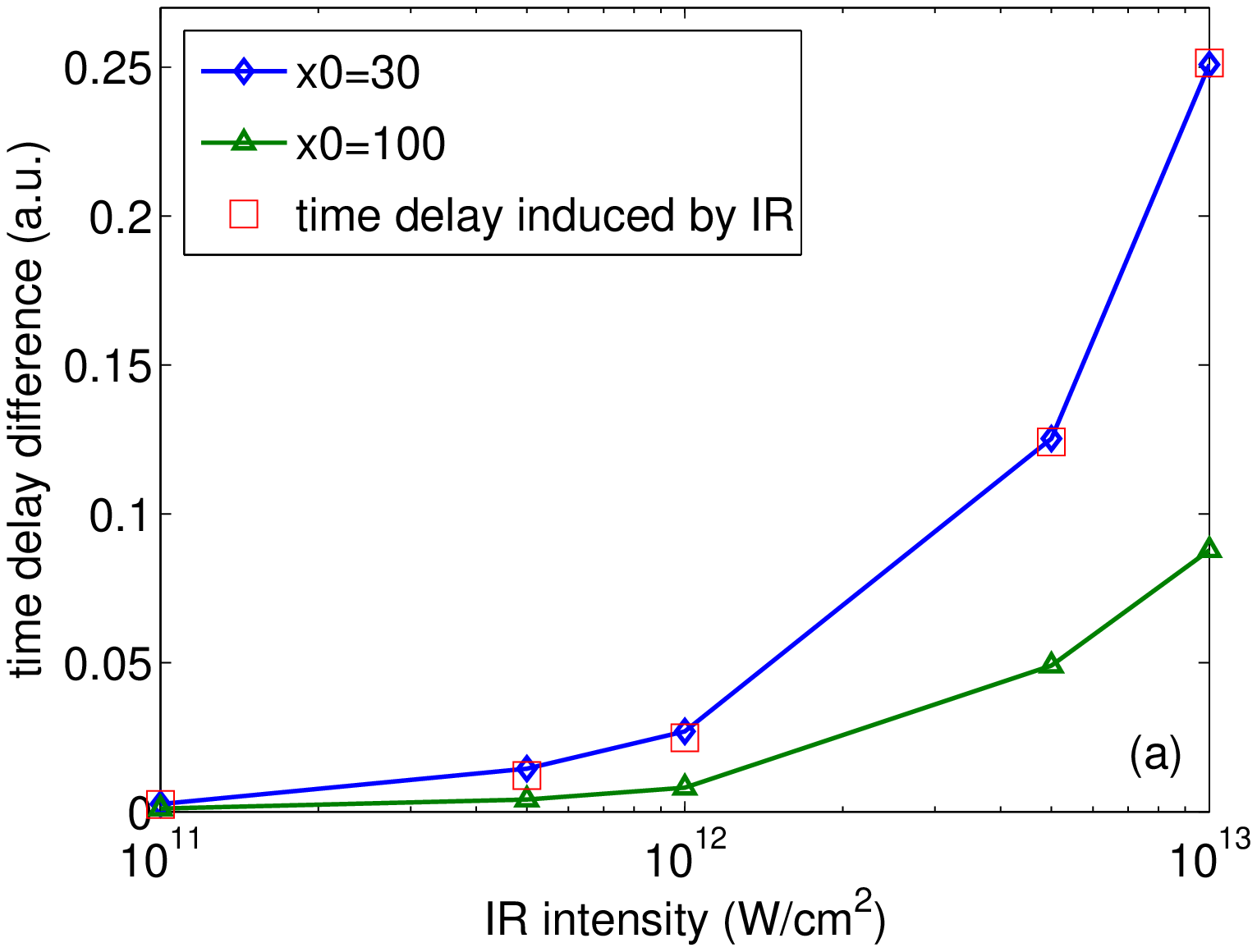}
  \includegraphics[scale=0.45]{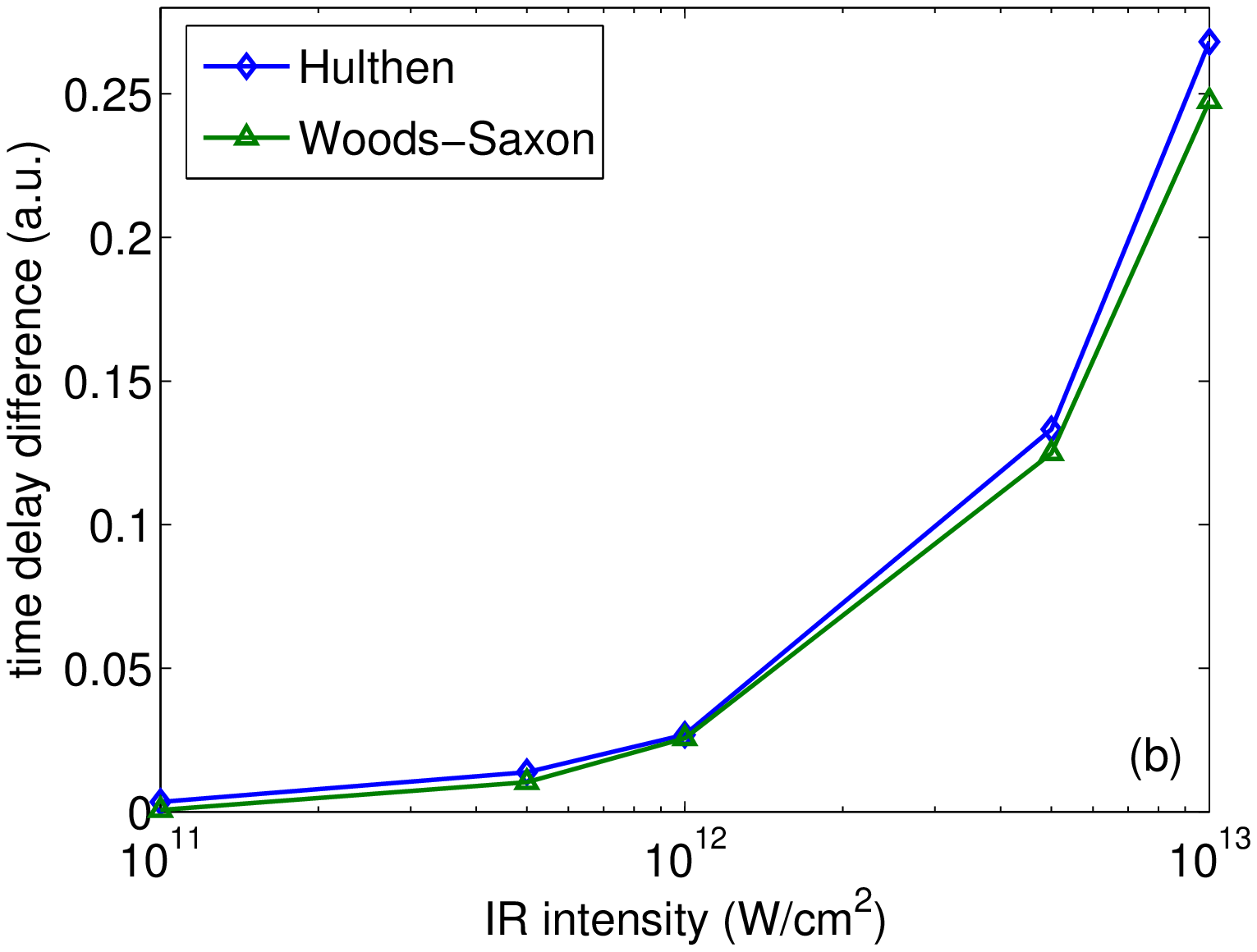}
  \end{center}
  \caption{
Time delay differences $\Delta T$ as a function of IR intensity: (a) the combined potential with $x_0=30$ (blue line with diamonds) and $x_0=100$ (green line with triangles), (b) Hulth\'en potential (blue line with diamonds) and Woods-Saxon potential (green line with triangles). The time delay $\Delta t_\text{IR}$ induced by the IR is shown as red open squares in panel (a). The XUV pulse is added at the middle (zero field, maximum vector potential) of the IR field. Laser parameters are: $I_\text{XUV}=1\times10^{15}$ W/cm$^2$, $\omega_\text{XUV}=100$ eV, $\tau_\text{XUV}=400$ as, $\phi_\text{XUV}=0$, $I_\text{NIR}=1\times10^{12}$ W/cm$^2$, $\lambda_\text{NIR}=800$ nm, $N_\text{NIR}=3$ cycle, $\phi_\text{NIR}=0$. (The color version of this figure is included in the online version of the journal.)
}
\label{delaydifference_IRintensity}
\end{figure}

In order to study the effect of the streaking field, we present in Fig.\ \ref{delaydifference_IRintensity}
the difference between the time delays in the streaking field to those without streaking field, i.e.\
\begin{equation}
\Delta T = \Delta t^{(\text{NIR})}_{{\Psi},R} - \Delta t_{{\Psi},R}.
\label{num_delay_difference}
\end{equation}
We obtained converged time delays for all short-range potentials once the NIR field ceased. In each of the present 
calculations the XUV pulse was applied at the center of the NIR field, which corresponds to the maximum of the vector potential.
Our results confirm our previous conclusions from studies using the Yukawa and Coulomb potentials \cite{su_submitted} 
that the time difference increases with increase of the intensity of the streaking field, but the effect remains small 
up to streaking field intensities of about $1 \times 10^{13}$ W/cm$^2$.

It is however interesting to note that the variation of the time delay with the streaking field intensity can be quite 
different for different short-range potentials. This can be seen, in particular, from the results 
for the combined Coulomb-Wood-Saxon potentials with different effective range, shown in Fig.\ \ref{delaydifference_IRintensity}(a).
We obtain a considerable smaller increase in the time delay as a function of intensity for the potential with a larger 
effective range ($x_0=100$, blue line with diamonds) than for the potential with shorter effective range 
($x_0=30$, green line with triangles).
This can be understood as follows. The effect of the NIR streaking field onto the time delay can be thought of as adding two 
contributions to the original delay $\Delta t_{{\Psi},R}$: a time delay caused by the IR field only ($\Delta t_\text{NIR}$) 
and a time delay induced by the coupling of the NIR field and the electrostatic potential ($\Delta t_\text{NIR-EP}$), i.e.\
\begin{equation}
\Delta t^{(\text{NIR})}_{{\Psi},R} = \Delta t_{{\Psi},R} + \Delta t_\text{NIR} + \Delta t_\text{NIR-EP}.
\label{delay_total}
\end{equation}
By combining Eq.\ (\ref{num_delay_difference}) and (\ref{delay_total}), we get for the time delay difference:
\begin{equation}
\Delta T = \Delta t_\text{NIR}+\Delta t_\text{NIR-EP}.
\label{num_delay_difference_new}
\end{equation}
In our numerical simulations we can determine $\Delta t_\text{NIR}$ by back-propagating the ionizing part of the wavefunction 
in the NIR field alone, determining the corresponding time and subtracting the time for the free particle back-propagation 
$t_{\Psi^{(0)},R}$. Since these test calculations in the back-propagation step do not account for the specific electrostatic 
potential and the ground state energies for each of the potentials considered are similar, our results for $\Delta t_\text{NIR}$
are almost the same for each of the potentials (for a given set of XUV and NIR streaking field parameters) and an example is shown 
as red open squares in Fig.\ \ref{delaydifference_IRintensity}(a). 

The time delay $\Delta t_\text{NIR}$ induced by the NIR field agrees very well with the full time delay difference $\Delta T$ 
for the combined potential with $x_0=30$ (blue line with diamonds in Fig. \ref{delaydifference_IRintensity}(a)), which indicates that the effect of a coupling between NIR 
streaking field and electrostatic potential is negligible in this case. This makes sense since we applied the XUV ionizing field 
at the maximum of the vector potential (and, hence, a zero of field) and the streaking field does not change significantly 
from zero while the ionizing wavepacket propagates within the electrostatic potential, as long as its effective range is small. 
As the effective range increases ($x_0 =100$) the coupling between electrostatic potential and (now varying) streaking field becomes 
stronger and the time delay difference decreases (green line with triangles in Fig. \ref{delaydifference_IRintensity}(a)).

\section{Conclusions}
We have applied a recently introduced numerical approach to obtain time delays in numerical simulations to XUV photoionization 
from different (short- and long-range) potentials. We have found, in general, excellent agreement with the results of calculations 
for the so-called WS time delay if the potential is short-ranged. In contrast, our numerical results confirm that for a 
long-range potential, such as the Coulomb potential, the time delay diverges in the limit of an infinitely large region and, 
hence, the WS time delay does not exist. Finally, we have found, in general, that the impact of a streaking field on the 
numerical results for the time delay is negligible as long as the intensity of the streaking field does not exceed 
$1\times 10^{13}$ W/cm$^2$, which corresponds to the intensity limit at which ionization of the target by the streaking field 
itself sets in.

\section*{Acknowledgment}
J.S. and A.B. acknowledge financial support by the U.S. Department of Energy,
Division of Chemical Sciences, Atomic, Molecular and Optical Sciences Program.
H.N. was supported via a grant from the U.S. National Science Foundation (Award No. PHY-0854918).
A.J.-B. acknowledges financial support by the U.S. National Science Foundation (Award No. PHY-1068706).
This work utilized the Janus supercomputer, which is supported by the U.S.\ National Science Foundation (award number CNS-0821794)
and the University of Colorado Boulder. The Janus supercomputer is a joint effort of the University of Colorado Boulder, the
University of Colorado Denver and the National Center for Atmospheric Research.

\end{document}